# THE FALLACY OF FAVORING GRADUAL REPLACEMENT MIND UPLOADING OVER SCAN-AND-COPY

**Keith B. Wiley, PhD[1] and Randal A. Koene, PhD[2]**

[1]Fellow, Brain Preservation Foundation
Author of *A Taxonomy and Metaphysics of Mind-Uploading*
kwiley@keithwiley.com
http://keithwiley.com, http://brainpreservation.org

[2]Founder, Carboncopies.org
randal.a.koene@carboncopies.org
http://carboncopies.org, http://minduploading.org

**Abstract**

Mind uploading speculation and debate often concludes that a procedure described as *gradual in-place replacement* preserves personal identity while a procedure described as *destructive scan-and-copy* produces some other identity in the target substrate such that personal identity is lost along with the biological brain. This paper demonstrates a chain of reasoning that establishes metaphysical equivalence between these two methods in terms of preserving personal identity.

## Introduction

Newcomers to mind uploading philosophy [Strout 1997] quickly become familiar with a shared set of canonical thought experiments. Two of the most popular are *gradual in-place replacement* and *destructive scan-and-copy*. The former consists of steadily replacing individual components of the brain, say neurons, with microscopic devices of functional equivalence, while the latter stabilizes the brain via vitrification or plastination, then sections and scans the static brain, and then instantiates the scan via whole brain emulation (WBE) in a computational substrate [Sandberg & Bostrom 2008, Koene & Deca 2014].

A popular intuitive point of view after considering these scenarios is to hold gradual in-place replacement more favorably than scan-and-copy with regard to successfully preserving personal identity (sometimes called consciousness, a shorthand for phenomenal consciousness, or our ongoing experience of ourselves) [Block 1995, Chalmers 2010, Olson 2010]. The argument essentially claims that gradual in-place replacement can achieve personal survival while scan-and-copy cannot. Gradual in-place replacement is granted the status of successful identity preservation while scan-and-copy is denigrated as producing a copy, the implication being that it embodies some other person, the person preceding the procedure having died [Corabi & Schneider 2012, Dorrier 2015, Hay 2014, Josh 2009, Morris 2013, Schneider 2014]. This view is often colloquially stated as "Even if scan-and-copy did work, it still wouldn't be *me*, just a copy." We restate this position in somewhat more formal terms in the *detractor claim*:

> Mind uploading via slow gradual in-place replacement will preserve the identity of the person associated with the biological brain and therefore represent their personal survival. However, mind uploading via destructive scan-and-copy, *even assuming technical efficacy* in that it produces exactly the same post-operative material brain as the gradual in-place replacement procedure, a brain of corresponding neural function and a person of corresponding psychological continuation, will nevertheless generate an entirely new identity, leaving the person from the biological brain "behind", and will therefore represent death of the person who preceded the procedure.

This article only addresses this precise detractor claim. It does not address questions such as whether either procedure may be impractical on technical grounds or whether certain considered procedures are more prone to imperfections or errors that would result in flawed duplications of neural function and corresponding psychological qualities. Note that the detractor claim is purely metaphysical in nature. Both the colloquial phrasing and the formal statement above grant total procedural success while nevertheless making differing claims about identity between two physical procedures. In this way, the claim can only be metaphysical, namely on the nature of *identity* (or alternatively *survival*, as we will discuss). Consequently, considerations of procedural conceivability, practicality, or flaws are outside the scope of this paper.

This article proposes that contrary to the opposing interpretations presented in the detractor claim, both scenarios should actually be considered metaphysically equivalent in the ultimate identity status of the minds they produce, either successful preservation of personal identity or failure thereby producing some other. To demonstrate this conclusion, we establish a transitive relation equating slow replacement with scan-and-copy using instantaneous replacement as an intermediary. This intermediate stage is similar to gradual replacement except that all neurons are replaced at the same time. We first demonstrate the equivalence of scan-and-copy with instantaneous replacement and then demonstrate the equivalence of instantaneous replacement with slow replacement. From this transitive relation, the conclusion is that it would be a fallacy to assign differing interpretations to any of the three scenarios.

Throughout this paper we refer to two alternative interpretations: the *preservation* of personal identity by the new substrate versus the emergence of some *other* identity. The terms *transfer* and *copy* sometimes indicate these two cases. *Transfer* implies associating one's personal identity with the new substrate. *Copy* implies metaphysical failure of that association, so that some *other* identity is created and associated with the new substrate instead; *copy* further implies that since the biological brain is destroyed by both in-place replacement and scan-and-copy, personal identity is presumed lost. The term *transfer* can be confusing when discussing mind uploading due to its overly spatial implications, so we prefer to speak of *preserving* personal identity in the new substrate. Similarly, the term *copy* is confusing when implying failed identity preservation because it is broadly used in other ways, e.g., copying a mind, personal identity, neural functionality, cognitive state (knowledge, memories, personality, etc.), brain scan data, physical brain structure or material pattern, etc. It appears in the popular label for the mind uploading method known as *scan-and-copy*, but there expresses purely technical aspects of data copying and does not speak to identity preservation one way or the other. For clarity, we avoid the term *copy* as used in the claim *it's-just-a-copy*, and instead consider whether personal identity is *preserved* when uploaded into a new substrate or whether some *other* identity emerges.

*Personal Identity and Personal Survival*

Sometimes it is useful to speak in terms of *personal identity* and other times in terms of *personal survival*. We must therefore disambiguate these two concepts. One could conceivably view them orthogonally, thereby offering four possible judgments to a destructive mind uploading procedure:

1. Identity is preserved and the person survives.
2. Identity is preserved but the person does not survive (i.e., they die).
3. Identity is not preserved, thereby producing some other identity, but the person survives anyway.
4. Identity is not preserved and the person does not survive.

Of the four possible judgments, the second one is fairly bizarre upon closer consideration, so we can dismiss it (what popular theory of mind or identity would work in this way?), but the other three are all possible. Crucially, the third option is possible, in which by some definitions, identity may be lost without sacrificing the notion of survival. This idea may seem equally bizarre at first glance, but actually it reflects more formal definitions of identity. For example, *mathematical identity* requires essentially identical states between compared objects, which would preclude saying that a middle-aged person has exactly the same identity as his childhood self. To accommodate this challenge, *personal identity* is generally loosened to allow sufficiently overlapping series of states, such as the slowly evolving psychological state of a person over his or her life, but this solution is problematic as well since a middle-aged person may overlap psychologically with his childhood self and his old-aged self even if his old-aged self has no remaining memories of his childhood. This is Thomas Reid's well-known *brave officer* problem and shows that identity cannot easily be considered transitive [Reid 1785]. Regardless of how we ultimately classify the thorny notion of *identity* across these scenarios, we readily interpret them as unambiguous *survival* and that is where the distinction between these two terms lies. Children don't conceptually *die* to beget their later adult selves, nor do adults represent survival only of their instantaneous selves but at the cost of the death of their former selves. We simply don't conceptualize survival in that way. So the third option above is at least tenable.

Another important distinction is that identity is often not viewed in an all-or-nothing way, but rather as being subject to partialities. This phrasing may seem at odds with the necessity that mathematical identity requires identical states, but as described above, this nuance is handled in the context of personal identity by considering partial overlaps in state. While we previously considered whether full identity might preserve across slow changes in state, here we consider whether identity itself conforms to ever-changing partialities that mirror the underlying changing state. If one allows partial identity, then a thirty-year-old shares some partially overlapping identity with his ten-year-old self, less so than his twenty-year-old self, but more so than his forty-year-old self. Furthermore, as shown with the brave officer presented above, his eighty-year-old self may represent absolutely no preserved identity from his childhood; the child's identity has been entirely lost, all the while *survival* has been entirely maintained.

We do not usually conceive that a person steadily loses survival (steadily dies) as his brain and mind undergo healthy neurological and psychological life events. Rather, we say a person has fully survived between any two moments arbitrarily chosen from his life's timeline, right up to the moment he dies. This usage also reflects how we speak of medical patients. A popular example in contemplations of mind and identity is a hemispherectomy, in which half of a person's brain is literally removed to treat epilepsy. Despite such a profound change in neurological (and some change in psychological) state, doctors and family nevertheless describe such a patient as having survived the procedure. Partial survival doesn't even enter into the conversation.

Taking nonbinary identity into consideration, the four cases above then become a continuous spectrum of identity possibilities with two binary survival labels available at any given point on the spectrum. So, identity might be judged to partially preserve via a destructive mind uploading procedure while at the same time the person is judged to either survive or die. The question central to this paper is whether we can rationalize assigning different identity/survival judgments to slow gradual in-place replacement and destructive scan-and-copy mind uploading procedures.

### *Identical Physical Product of Mind Uploading Procedures*

When contemplating scan-and-copy, we assume that the product of the procedure will not merely be a software simulation of neurological behavior running on an otherwise semi-conventional computer, but rather that the final product is, in fact, the *exact same* product that is produced by an in-place replacement procedure, down to the last atom. The functional theory of brain and mind on which whole brain emulation and mind uploading are predicated does not require this property; a simulation should suffice. However, comparison of the thought experiments this paper investigates benefits from the removal of unnecessary differences between the various upload products in question. If possible, we should consider scenarios that all produce the *exact same* physical result, but merely by different means, and then investigate their metaphysical differences, if any. In-place replacement implicitly produces a device that is unlike any computer that has ever existed before, namely a system comprising billions of microscopic neural prosthetic devices (artificial neurons) physically networked with a topology (i.e., a connectome) that reflects the topology of the biological brain it replaced. Recent advances in neuromorphic computing approximate what this futuristic technology may resemble [Monroe 2014]. Scan-and-copy offers more diverse options, such as pure simulations, but conceivably it could proceed by instantiating the scan data in a three dimensional nonbiological replica of the original brain, i.e., precisely the same object that in-place replacement would have produced. The fundamental question is, can we reasonably conclude that these two physical objects, identical in structure yet arising from different procedures, should receive any differing status in terms of identity preservation or personal survival, or is such a conclusion unwarranted?

### *Reduplication*

One extrapolation of mind uploading thought experiments that seems to occur to virtually everyone is the problem of *reduplication*, in which the recreation of a brain's salient properties (and thereby a second instance of its mind) is accomplished nondestructively, so that the original biological brain and mind remain intact. This line of reasoning can just as easily apply to biological scenarios (clones from throughout science fiction lore) as to mind uploading scenarios in which the new system is of an ostensibly computerized form.

That fact that reduplication is easier to envision in some scenarios, like scan-and-copy, than in others, like in-place replacement, is often used as an argument that identity preservation stands on weaker ground in the former case than in the latter. In fact, the mere notion that reduplication can conceivably occur in scan-and-copy is often taken as proof, in itself, that identity preservation is simply off the table for *any* version of scan-and-copy, including the destructive variant. After all, why should the identity status of an upload be affected by the disconnected and unrelated original brain, especially when the original brain may or may not still exist depending on the circumstances? However, there are commonly under-appreciated extrapolations of in-place replacement that show it needn't necessarily be destructive (see the scenarios under 1.1.2 in the taxonomy in Wiley [2014]). Consequently, the mere *possibility* that a given procedure might be altered to operate nondestructively appears to offer no insight at all into the question of identity preservation.

Since this paper explicitly makes a comparison to a destructive procedure (gradual in-place replacement in its popular form), we dispense with additionally analyzing nondestructive procedures of either sort in the interests of conciseness. To refuse to even entertain reduplicated scenarios may come across as a mere avoidance of the issue, but reduplication is such a rich area of inquiry as to justify its own investigation as a central topic, not a tangential one, as is done in other writing [Cerullo 2015, Wiley 2014]. Therefore, in this paper we focus our *comparison* on destructive scan-and-copy to confine the discussion to the popular example of destructive in-place replacement and the most similar variant of scan-and-copy possible.

In any case, as to whether reduplication should undermine our argument, Parfit [1984] (and others) have shown that this needn't be so. One popular thought experiment considers hypothetical hemispherectomies in which the left and right hemispheres of the brain are removed and then continue to live in separate bodies. A further extrapolation even clones (or alternatively, uploads) the missing hemisphere in each result to reform two complete brains [Wiley 2014]. The question asks: what happened to the original identity? A complete reflection on such cases (which cannot fit here, but see Parfit [1984]) shows that reduplication leaves the question of identity unresolved. Rather than prove any claim that identity favors the original brain, reduplication merely confirms what we see as unjustified, yet deeply held *prejudices* favoring in-place procedures, prejudices which don't withstand closer scrutiny. As stated, tackling such prejudices is addressed in Cerullo [2015] and Wiley [2014]. We will briefly revisit the reduplication issue in the conclusion, but otherwise will not entertain the notion again in this paper.

### *Style of the Argument*

This paper first considers a spectrum of spatial displacements between a biological brain and its replica resulting from a mind uploading procedure, and second considers a spectrum of temporal replacement rates at which an in-place replacement procedure could conceivably be performed. Similar thought experiments have been presented before. As we will discuss, Chalmers and others have described the very same temporal spectrum before, but those past descriptions have not carried through on the exploration by extending from the temporal spectrum to the spatial spectrum, thereby leaving half the argument unstated. This paper shows that these two examples are actually intimately linked, representing two halves connected end-to-end to form a single overarching chain of reasoning. This chain is characterized by the transitive relation we will present, tying scan-and-copy to instantaneous in-place replacement, and then on through to slow in-place replacement.

Another spectrum that has been offered in past literature is Parfit's *psychological spectrum*, taken from his discussion of Williams' thought experiment in which one person's memories are intermingled with another's (but without substituting any physical material in the brain) such that extreme points on the spectrum indicate two completely distinct memory sets (as with any unrelated pair of people) and intermediate points indicate the loss of some memories from each of the two people, but the inclusion of memories from the other (the similarity to the brave officer is inescapable). Later, Parfit proposes yet another spectrum, the *physical spectrum*, in which endpoints indicate a person before and after total material replacement and intermediate points indicate steps along the way by replacing some portion of cells in the body (and later still Parfit combines both spectrums into a third spectrum). The physical spectrum obviously closely mirrors the *process* itself of in-place replacement mind uploading, in which neurons are incrementally replaced, although in mind uploading speculation it is more common to confine considerations to neuronal replacement as opposed to full body cellular replacement (a distinction that is not too important here).

Parfit uses the psychological and physical spectrums to investigate whether identity is closer to a psychological or physical concept [Parfit 1984]. In the case of the psychological spectrum, but by preserving the same physical material across the spectrum, should the two people at the ends be identified identically or differently? Similarly, in the physical case, in which psychological properties hold across the spectrum but all physical material is steadily replaced, should the ends be identified identically or differently? Parfit investigates how we should quantify the transitions along these two spectrums. Sharply, with identity flipping at some point along the spectrum? Smoothly, with identity being some sort of blend? Or as Parfit concludes, "emptily" meaning that it is not really a reasonable question to attempt to label the intermediate points, and that there should not necessarily be an answer at all (or at least not one we can hope to discern). Parfit leaves these questions mostly open as he ventures into his later discussion, in which he presents his notion of *Relation R*, which is his proposed alternative to psychological and physical identity.

In this paper we consider two other spectrums, as summarized above. Similar questions arise. Should the endpoints be identified identically or differently? Or in terms of *survival* instead of *identity*, should traversing either of these spectrums indicate a transition from survival to death? Furthermore, as in Parfit and Williams' examples, should intermediate points along the spectrums be categorized sharply, smoothly, or as Parfit concluded, is it an empty question to inquire about the status of intermediate points? This paper also considers an end-to-end combination of the temporal and spatial spectrums in which slow in-place replacement resides at one end and scan-and-copy resides at the other (with instantaneous replacement residing at the midpoint). The detractor claim states that we should emphatically assign different labels to the endpoints of this two-part spectrum, with one end indicating identity preservation and survival and the other indicating a totally different identity and/or death. The intent of this paper is to show the fallacy of such reasoning.

## Equating Destructive Scan-and-Copy with Instantaneous Replacement

We begin by showing the equivalence of scan-and-copy and instantaneous replacement. In the former, the brain is frozen, sectioned, and scanned, and the scan data is used to build an artificial brain composed of billions of prosthetic neurons of perhaps electronic or optical design (or even some more exotic mechanism yet to be developed). In the latter, the brain is infused with billions of prosthetic neurons that travel to assigned biological neurons in a one-to-one relationship, passively observe the neurons to model their behavior (their functions), and then instantaneously replace all the neurons at the flip of a master switch. For the purpose of the thought experiment, these two procedures produce the exact same physical product, as if the scan-and-copy method had frozen the brain for subsequent sectioning at the very same moment in time that the instantaneous replacement procedure would have otherwise flipped the master switch.

Bear in mind that we are not asking questions about practical utility, such as whether either of these procedures is overly difficult to perform. One might challenge that instantaneous replacement as described sounds veritably implausible from a practical standpoint, but that is irrelevant to our considerations. Much as Einstein could ponder about riding beams of light and gazing into mirrors along the way, we can consider technically unlikely scenarios for their purely philosophical implications. More to the point, the detractor claim explicitly grants us this leeway in its own allowance of technical efficacy, all the while denying metaphysical equivalence. So, for that sake of metaphysical inquiry, we assume that these procedures work; that is, they produce a nonbiological brain whose neural operation is a natural continuation of the biological brain and whose associated mind is a natural continuation of the original psychological properties as well.

At this stage of the analysis, the question is not whether identity is preserved or whether the person survives. Those questions will be addressed later. The sole point in this section is that there isn't a reasonable distinction between destructive scan-and-copy and instantaneous in-place replacement. Whatever metaphysical result in terms of identity or survival we assign to either of these procedures should also be assigned to the other. They are not only physically identical in their result, they are metaphysically identical as well.

### *Spatial Translation*

Some readers may feel that the distinction between instantaneous replacement and scan-and-copy lies in the likelihood that scan-and-copy involves a greater spatial translation of neural function from the biological brain to the new substrate, perhaps even on the belief that any in-place replacement procedure requires no spatial translation at all. In these debates it is sometimes asked how function and identity could possibly *move through space* from one brain to another. For example, Corabi and Schneider state:

> Consider first that the mechanism by which the person (i.e., the propertied substratum) would move instantaneously from the brain to the computer is problematic, even on the assumption that only a short distance needs to be traversed. Not only does this involve an unprecedentedly rapid kind of motion for a person to follow, but this sort of motion is oddly discontinuous. For it is not as though the person moves, little by little, to the computer, so that a step-by-step spatial transition from brain to computer can be traced [Corabi & Schneider 2012].

Descriptions like the one above exemplify a conceptualization of minds as having genuine *spatial locations* (e.g., inside heads), such that uploading must consist of a true motion through space by the purported mind entity. To state the alternative, we favor the notion that minds are entirely nonphysical and nonspatial, and are merely *instantiated* by physical brains. Being nonphysical in nature, minds are not so much spatially colocated with brains as they are associated with brains, since they are not capable of spatial locations to begin with. If such wording strikes the reader as dualistic, bear in mind that it implies property dualism at most, certainly not substance dualism.

On the apparent assumption that minds and identity have true spatial locations, it may seem that minimizing or even negating any spatial translation of the neural functionality and metaphysical mind implies greater compatibility with identity preservation, especially as contrasted with spatially transferring a person's mind and identity across a room to a cloned or robotic body. However, we will show that this position is difficult to rationalize.

First of all, the possible assumption that in-place replacement involves no spatial translation is simply incorrect. While scenario 1.1.1.1.2 in the taxonomy presented in Wiley [2014] offers a detailed explanation, the following summary is straightforward. There is always some spatial translation of function as a neuron *hands off* its neural behavioral role to a nearby microscopic device, and furthermore, the hand-off occurs discontinuously through space (as described by Corabi and Schneider above); this functionality does not move smoothly through space from location A to location B (in so far as we can speak of nonphysical functionality residing at a location in space or moving through it in the first place! Others would seem to agree [Hopkins 2012].). Rather, at one moment, some neural function is being performed by a biological neuron in one location; then at some later time that same function is being performed by a nearby artificial neuron in a spatially discontinuous and different location and the biological neuron is no longer operational. The two entities, a neuron and its replicant, cannot spatially coincide even if they are so near one another as to physically abut (~10–100 microns from center of neuron to center of replicant)—there is simply always some distance to be considered (to a committed functionalist, this consideration of the nuances of physical distance may sound tedious and irrational, for what possible relationship could there be between physical distance and nonphysical function, but we must consider such matters to complete the argument).

Recognizing that some spatial translation is always involved, we can consider whether there remains a fundamental metaphysical difference in this translation between the two scenarios. In scan-and-copy, the distance over which this functionality purportedly transfers is on the order of several centimeters or a few meters at a minimum (from one body to another in the same room) or arbitrarily further. For example, teleportation, a variant of scan-and-copy, could transfer mental function any conceivable distance, including across the universe and/or far into the future. But as shown above, in-place replacement involves some spatial translation as well. Consequently, any spatial distinction must now hinge on distance comparisons. Curiously, Wiley [2014] shows that the micron-scale per-neuron translations that occur during in-place replacement accumulate hundreds or even thousands of kilometers of total discontinuous spatial translation of function, despite the possible presumption that no spatial translation is involved. Once we have already accepted spatial transfer of function on the order of hundreds of kilometers—albeit distributed over billions of neurons—what difference does a few meters across a room really make?

The fact of some nonzero spatial translation in both procedures leaves us with two possible conclusions. We can declare that translations below some maximum permitted distance have different results from farther translations, perhaps that they are tolerable as a true preservation of identity while farther translations are deemed preservation failures (and furthermore invoke another mind out of nothing where at shorter distances, perplexingly, no other is created at all!), or alternatively, we can accept translations of any distance as being functionally and metaphysically equivalent. The problem is how *identity preservation* can either suddenly change or alternatively blend into *other identity*. A blend would take the form of a smooth transition from preserved identity to invoked identity relative to the distance translated from the biological brain to the new substrate, with intermediate distances indicating some combination of two identities, part of the original and part of the new. Both proposals, cutoff and blend, face serious challenges. Bear in mind that on both sides of a sharp identity cutoff relative to translation distance, and across any proposed spectrum of translations corresponding to a blend, the exact same physical result occurs in all cases, be it instantaneous replacement with translations of microns, teleportation with translations of kilometers, or scan-and-copy with translations of centimeters or meters within an operating room. The following questions then arise:

1. How could one rationally assign a sharp cutoff in translation distance where identity preservation suddenly flips to new identity invocation?
2. Alternatively, how can one rationally conceive of a blend, considering the following issues:
   a. What is the range within which the blending function occurs, but beyond which the resulting identity is entirely new and the original is entirely lost? Can such an assignment be theorized in a nonarbitrary manner?
   b. How can it be rationally defended that various identical physical results, differing by not one atom in their material composition, but only in their locations relative to the biological brain, house metaphysically different blends of identity, perhaps 1%-original:99%-new, 30%:70%, 50%:50%, or 80%:20%, etc.?
   c. While we might just barely conceive of partial identity preservation or survival (although this notion really only makes sense when contemplating imperfect procedures, which is not on topic!), how can we possibly conceive of partial *invocation* of a brand new identity? What does it mean to bring into existence only a fraction of a new identity to then blend with the partially preserved original?
   d. Ultimately, the issue of blending asks: why (and by what physical mechanism) should a procedure in which the replica brain resides closer to the biological brain allow the replica to receive more preserved identity than an identical procedure in which the replica resides farther away, and contrarily, why should a closer replica invoke less brand new identity and a more distant replica invoke more brand new identity?

The proposal of a cutoff or smooth blend in spatial translation requires that some metaphysical property of personal identity can tolerate spatially discontinuous translation over a distance of tens of microns, but seemingly not over a distance of centimeters, meters or perhaps kilometers—or light-years (including vast accumulations of distance gathered from billions of otherwise micron-scale translations in the case of in-place replacement, as described above). There is no evidence that such a metaphysical property of consciousness and personal identity survival exists that is subject to these seemingly bizarre physical and spatial effects, especially considering that we are assuming perfect technical efficacy (and, in fact, identical physical results) across any such spectrum of spatial translations.

### *Conclusion of the Spatial Argument*

The most rational conclusion to the questions that arise when considering spatial translations of identity is that identity and survival simply have no relationship to the spatial distance between the biological brain and the prosthetic replacement. In fact, we advocate for the stance that function, mind, and identity don't even have spatial locations to begin with and that any concern or challenge about their alleged motion through space is a misnomer. In this way, scan-and-copy and instantaneous replacement are then judged identically in terms of identity preservation or survival. They either both represent full preservation and survival, or they both represent full newly invoked identity and total lack of survival, i.e., death.

Bear in mind that the counterproposal that the distinction lies in piecewise replacement as opposed to distance is irrelevant in this stage of the analysis since the comparison is being made against global instantaneous replacement, in which all neurons are simultaneously replaced by their artificial prosthetics. We address piecewise replacement in the next section.

We should briefly address the possible counterargument of a Sorites paradox (aka, the paradox of the heap), in which by initially considering a heap of sand, and by then removing individual grains without seemingly changing classification, we ultimately conclude that a single remaining grain of sand constitutes a genuine heap (hence the paradox). The distinction is that in the case of heaps of sand we can look at the two endpoints of the spectrum (a large mound of sand and a single grain of sand) and clearly label them differently with regard to the term *heap*; one endpoint unambiguously does not conform to the definition of a heap. Put another way, the fact that there is a difference between the endpoints is not in dispute in a Sorites paradox; it is a *premise* of the paradox. The Sorites paradox lies entirely in the apparent difficulty of determining where or how the indisputable difference occurs across a spectrum of seemingly inconsequential transitions. The situation is completely different on the question of metaphysical identity or survival across a spectrum of mind uploading procedures that all produce the exact same physical result. The question is not merely where or how along the spectrum a supposed transition in identity or survival occurs, but more fundamentally, *whether* there is a difference between the endpoints to begin with! Concluding that there is no difference simply erases the premise of any purported Sorites paradox. So long as the subject of debate is not only the nature of the transition but also whether the endpoints even differ in the first place, any claim of a Sorites paradox is putting the cart before the horse.

So long as the complex issues and questions shown above remain open, the default philosophical stance should be the simpler theory, namely to entirely dispense with the notion that identity is dependent on translation distance (Occam had some choice thoughts on such matters). Furthermore, and this is by no means a negligible point even though we are appending it at the end here, doesn't it seem remarkably suspicious that in such pondering we are tempted to assign a cutoff such that distances we can't easily distinguish in our daily experience (microns) are considered co-identified and safe while distances we casually comprehend as distinct translations (meters, kilometers, etc.) are considered identifiably separated and unsafe? Such a designation smacks of anthropocentric reasoning. Would aliens who are a few orders of magnitude smaller or larger than humans assign this cutoff differently?

## Equating Slow Replacement with Instantaneous Replacement

In the previous section we used an argument in the spatial domain to establish the equivalence of destructive scan-and-copy and instantaneous in-place replacement. We described a spectrum of spatial displacements across which one could conceive of a destructive procedure, with instantaneous replacement residing toward the short displacement end and scan-and-copy residing at a farther displacement. We clarified the possible misunderstanding that one might expect the spectrum to start at zero displacement for instantaneous replacement, showing that, in fact, instantaneous replacement corresponds to a displacement in the range of tens of microns. This clarification obviated any challenge that the crucial distinction might lie in the dichotomy between *any* displacement versus *no* displacement. We then exposed the wide array of problems that arise from associating identity with spatial displacement (or even spatial locations at all) and finally concluded that all procedures along such a spectrum should be judged as metaphysically equivalent with regard to identity preservation and survival.

In this section we use a similar argument in the temporal domain to establish the equivalence of instantaneous replacement and slow replacement. As with spatial displacement, any purported distinction between various rates of replacement proves to be highly problematic, and furthermore, to be metaphysically unmotivated in that (exactly as was shown for spatial considerations) there is no particular reason to initially believe identity bears a relationship to replacement rate in the first place. Consequently, until shown otherwise, the more parsimonious and more likely conclusion is that there simply is no such relationship.

### *Replacement Rate*

Variants of the temporal argument have been proposed before, such as Chalmers' chapter in Blackford & Broderick [2014]. What we offer in this paper is a previously unstated connection between the spatial and temporal examples, thus closing a transitive relation from slow replacement to instantaneous replacement and then to scan-and-copy. The temporal argument considers a spectrum of temporal rates of gradual replacement, measured in neurons replaced per second or some comparable metric. Scenario 1.1.1.1.1 of the taxonomy in Wiley [2014] offers a detailed description, which incidentally also shows that initial expectations that slow in-place replacement is positively languid may not agree with replacement rates required for a procedure of reasonable expediency. For example, a total procedure time of 100 days, operating around the clock, would require replacing 10,000 neurons every second nonstop. To complete the procedure in a twenty-four hour period would require continually replacing one million neurons every second for an entire day. Putting aside whether a procedure of practical duration would actually be meaningfully *slow*, incremental replacement is nevertheless how gradual replacement is generally presented and in this way it differs from scan-and-copy, which equates to global instantaneous (or at least discontinuous) replacement. At faster replacement rates, slow replacement approaches instantaneous replacement, thereby resembling scan-and-copy in some crucial aspects.

Some readers may be inclined to assume that the fast end of the spectrum would lead to failed identity preservation or failed survival, with some other person resulting from the upload process, while the one who entered the procedure died. In fact, anyone conforming to the detractor claim should now deny identity preservation to instantaneous replacement due to its equivalence with scan-and-copy, as shown in the previous section. One reason why slow in-place replacement is often favored is that it might maintain some notion of a *stream of conscious continuity* [Chalmers 2010, Van Gulick 2014]). In debates about the detractor claim, arguments to the effect that nongradual procedures fail to preserve this stream are often raised, even when allowing for the assumption of technical efficacy, i.e., total procedural success in capturing neural and consequent psychological qualities. However, defending such a conclusion falls to its proponent as it is utterly metaphysically speculative—perhaps the alleged conscious stream can just as easily survive instantaneous replacement (or scan-and-copy) after all, or perhaps the entire idea of a consciousness stream is simply flawed [Blackmore 2002, Dennett 1991]). For example, by what theory of a metaphysical stream of consciousness would this stream be contingently preserved across various physical procedures differing only in replacement rate, yet producing identical physical products? Remember, we aren't deliberating over whether such a consciousness stream might genuinely fail if the procedure itself is *physically* flawed at faster replacement rates so as to produce an imperfect nonbiological brain and associated mind. The detractor claim entertains absolute procedural success even while denying preservation of the critical metaphysical properties, so breaking the consciousness stream (if it even exists in the first place) would have to occur across a spectrum of procedures that are invariant in their final physical products. One can certainly propose the nascent idea that two procedures that yield physically identical results nevertheless differ in the preservation of an alleged metaphysical stream of consciousness, but that is a claim requiring significant substantiation (which is currently lacking) and should not be held as anything exceeding pure conjecture until otherwise shown.

Aside from notions of a stream of consciousness, to otherwise hold the position that slow replacement preserves identity but that instantaneous replacement does not implies that some success-to-failure transition must happen somewhere in the range of possible replacement rates. One might claim that there exists a cutoff along the replacement rate spectrum where personal identity preservation suddenly flips to producing some other identity (despite no change in the physical product). Immediately below the cutoff, the rate would be considered sufficiently slow to allow total preservation of personal identity, yet at an infinitesimally faster rate, the interpretation would be to utterly reject preservation, resulting in some other identity instead. While all mental functionality would still technically instantiate in the new substrate (throughout this discussion, all considered procedures produce the exact same physical result), thereby producing an identically functioning, thinking and behaving person (Chalmers' *functional isomorph* [Chalmers 2010]), detractors would apparently claim that the crucial *properties of personal* *identity* have failed to ride along, and for no better reason than the minutest increase in replacement rate. Proposing such a cutoff requires not only a defense that it is a valid concept in the first place, but furthermore some notion of where on the replacement rate spectrum it actually resides.

Some readers might propose that the transition from identity preservation to other identity not necessarily conform to a discrete cutoff, but rather to a smooth continuum cross-dissolving from full preservation (slow replacement) to full other (instantaneous replacement) as the replacement rate increases, and with intermediate replacement rates resulting in a mixture of both. The proposal of blends of partially preserved and partially created identity suffers nearly identical problems to those faced when considering the spatial spectrum of displacements. It claims that given two hypothetical nonbiological brains, materially identical to one another, but which resulted from replacement procedures spanning longer and shorter times respectively, we should conclude that the former upload now houses a greater proportion of the original identity than the latter (and by implication houses a smaller proportion of an invoked new identity). This conclusion faces the same problems listed above, briefly summarized from that list as (2a) how to concisely define the blend as a function of replacement rate, (2b) how various materially identical results can house different proportions of two distinct metaphysical identities, (2c) how to conceive of invoking some *incomplete fraction* of a new identity, and (2d) why, short of an unsubstantiated hunch, a procedure spanning a longer time should necessarily receive a greater proportion of preserved identity. The conclusion we draw is the same as in the spatial case: there simply is no particular reason to propose a relationship between metaphysical concepts like identity and physical temporal concepts like the rate at which a medical or technical procedure is conducted (assuming it produces an identical physical product regardless of replacement rate). One can posit such an abstract and utterly metaphysical relationship, but without an adequate description of such a relationship, we should consider it to be no more than nascent supposition (and quite likely unfalsifiable as well).

One might counter that both theories are equally unsubstantiated and therefore that any preference is arbitrary: that it is no more speculative to propose a connection between metaphysical identity and the speed of some physical procedure (which produces an identical physical product regardless of speed) than it is to propose that no such connection exists. However, this is precisely the sort of thing that Occam's Razor addresses: parsimony. One of these theories proposes an *additional* metaphysical property, crucially with no additional physical consequences, while the other proposes no such property. The two theories are identical in their physical implications since neither implies any difference in the physical product of uploading (the resulting nonbiological brain cannot even theoretically reveal a tangible difference in physical measurement on the basis of this purported metaphysical connection). But, one of these theories is more parsimonious by dispensing with a superfluous (and undetectable) metaphysical-identity-to-replacement-rate connection for which there is no apparent (or even conceivable) evidence. We previously described the proposed metaphysical relationship as unfalsifiable. By this we mean that even if such a metaphysical relationship did tie identity to concepts like replacement rate, there would be no way to ever detect or verify it. Consequently, it can have no *discernible* bearing on our model of reality. To claim that the theory of such a metaphysical property stands on even ground with the theory of no such property requires a description of how this metaphysical connection could ever be detected or verified.

To clarify a possible point of confusion, we are not considering whether there are valid practical or strategic reasons to choose one procedure over another, nor are we considering whether gradual in-place replacement has a physiological and technical speed limit, i.e., whether replacement above a certain rate might yield an upload that fails to adequately preserve the original neural functionality and cognitive behavior and consequently fails to sufficiently resemble the person who entered the procedure. While the matter of a technical speed limit is doubtlessly crucial from an engineering perspective, it is irrelevant to this philosophical discussion *precisely* because detractors often grant this concession when denying identity status to an uploaded individual. The spirit of the detractor claim is that *even if perfect* *preservation of function is achieved (and even if the resulting physical product is identical across the considered scenarios!)*, we should nevertheless regard the person preceeding certain procedures as dead and refuse to grant the uploaded mind the same personal identity. In this article, we have only considered whether *metaphysical identity* is subject to some transition from preserved to other personal identity (even when the physical end product does not differ) along spectrums of spatial displacements and temporal replacement rates. Technical efficacy is off the table, not due to our restriction, but due to the widely held detractor claim itself.

*Conclusion of the Temporal Argument*

Since we are assuming identical end products across all scenarios, a distinction in metaphysical status between scenarios like slow gradual in-place replacement and destructive scan-and-copy seems the unlikely state of affairs, regardless of which status is ultimately assigned. Additionally, as to that status, since we are assuming technical efficacy in the duplication of both neural function and psychological states, a transition of identity from preservation to other seems unlikely as well. Until a sufficient theory is presented as to why such a metaphysical state-change should be expected, the default position should be that no such state-change occurs, either spatially or temporally.

If we conclude that no transition along the temporal spectrum of replacement rates is ultimately defensible, then we should grant identity preservation status to both a slowly replaced and an instantaneously replaced individual. Furthermore, given the established equivalence of instantaneous replacement and scan-and-copy, we should then extend the same status to scan-and-copy as well. This reasoning results in a conundrum for the position that slow in-place replacement and scan-and-copy should receive differing identity or survival statuses. One of the detractor's positions should dominate the other and resolve the fallacy. For example, if a detractor holds to the position that slow in-place replacement preserves identity, then he or she should accept the identity of instantaneous replacement and scan-and-copy as well. Alternatively, if a detractor holds that scan-and-copy does not preserve identity, then he or she should accept that neither instantaneous nor slow in-place replacement can do so either. The realization of this inconsistency is the crux of this article.

## Conclusion

If we disregard the claim that identity preservation becomes other identity somewhere along the spatial spectrum, then we conclude that instantaneous replacement and scan-and-copy are functionally equivalent. If both the temporal and spatial distinctions fall, thereby equating slow replacement and instantaneous replacement, then slow in-place replacement is also functionally equivalent to scan-and-copy. This conclusion does not prove that scan-and-copy must be regarded as a successful preservation of personal identity or a form of personal survival, but it does demonstrate that both procedures should be judged in the same manner: we either grant scan-and-copy successful status and stop denigrating it as a mere copy lacking in proper identity status, or we refuse to grant identity status to the oft-favored slow in-place replacement and deem both procedures to be metaphysical impossibilities.

As indicated toward the end of the previous section, we urge that the ultimate conclusion for both procedures should be identity preservation and/or personal survival, as opposed to invoked other identity and/or death. The general notion of identity appears to tolerate piecewise replacement, spatial translations, and even whole parcel scan-and-copy. While the classic thought experiment of The Ship of Theseus merely poses as an open question whether identity can survive piecewise replacement, many conceptually identical examples are taken for granted to do so, such as our own bodies, in which very little matter persists over the long term, buildings under lifelong renovation, waves in any physical medium, colonies, etc. With regard to scan-and-copy, multiple copies of a book or recordings of a song are casually regarded as multiple physical instantiations of a singleton information pattern (even in this very sentence we referred to the book and the song in singular vernacular, and all readers, including detractors on our central issue, accepted that phrasing without even noticing the irony). In other words, we apply the word *copy* to the physical instance of a book (a token) while recognizing that the underlying physically embedded information (a type) has neither *duplicated* nor even really *changed location* in a meaningful sense; those are properties of the physical instantiation, not the associated information. We can easily admit that various physical instances represent different tokens of a single common type. In this way, and confining our considerations to destructive procedures, multiple tokens not overlapping in time (the biological brain before any uploading procedure begins, and the nonbiological brain after the procedure is completed) can rightly be judged to instantiate the same fundamental type, i.e., *identity*, and by implication to further indicate *survival* of that identity across time and between tokens. All of these examples suggest that we should not only grant the same status in all the scenarios this article has considered, but specifically a status of preserved personal identity.

One final thought on this matter involves the paradox of allowing preservation of identity in a scan-and-copy scenario that may not require destroying the biological brain and mind (often known as *nondestructive* scan-and-copy, or *reduplication* in philosophy of mind). This possibility was only briefly mentioned in the introduction. Some readers, and most detractors on the question of preservation status in scan-and-copy scenarios, take this paradox as proof ipso facto that identity preservation is simply beyond the pale in any scan-and-copy scenario. When considering such a dismissal, one should first realize that scenarios like those under 1.1.2 in the taxonomy in Wiley [2014] show that in-place replacement can also present the duplication challenge wherein the biological brain remains unharmed while a duplicate substrate is produced. This under-appreciated possibility of a nondestructive *in-place* procedure demonstrates that the mere notion, in and of itself, of nondestructive scan-and-copy implies nothing whatsoever as to whether the identity status of scan-and-copy should be summarily dismissed. Any judgment must rely on other factors. We propose that the best solution to the paradox of mind uploading in which the biological brain survives is to adopt a completely different third model of identity that conforms to neither the transfer (preservation) nor the copy (other) interpretation, but rather to the notion that minds and personal identity can conceptually split (also known in the literature as branching or fission) into multiple descendants of equal primacy to their common ancestral mind and identity. A thorough description of this recommended identity model will not fit here, but please see Cerullo [2015] or Wiley [2014] for detailed descriptions.

## Acknowledgements

We would like to thank Michael Cerullo, Alexander McLin, and Oge Nnadi for their collaboration on this topic and on refining and editing this paper.